# Large Eddy Simulations of turbulent convective flow through a periodic groove channel

Auronil Mukherjee[1] and Arnab Chakraborty[2, *]

[1]Department of Applied Mechanics, IIT Madras, Chennai-600036, Tamil Nadu, India

[2]Department of Mechanical Engineering, IIT Madras, Chennai-600036, Tamil Nadu, India

**ABSTRACT**

*The use of extended surfaces find wide range of applications in heat transfer devices for achieving heat transfer augmentation like gas turbine blade cooling and nuclear reactor core since the last few decades. So, understanding the underlying flow physics physics and the transport phenomenon governing the heat transfer enhancement is the goal of the present study. In the present investigation, numerical computations of turbulent forced convection through a periodic groove channel are carried out using large eddy simulations. The lower wall of the grooved channel is provided with constant heat flux while upper wall insulated. Computations were carried out using WMLES model in LES formulation implemented in a finite volume based commercial software ANSYS Fluent 19.2®. The simulations are performed over varying Reynolds numbers (Re) range of 3000-30000 at different ratios groove width to channel height (B/H) in the range 0.75-1.75. The groove pitch ratio, and depth ratio kept constant of magnitude 2 and 0.5 respectively. Estimation of coefficient of heat transfer, associated frictional losses, and magnitude of heat enhancement are systematically carried out and compared with reported results in literature obtained using RANS framework reported in literature. The results obtained using LES show improvements in the heat transfer rate by a reasonable magnitude of 45% while the associated frictional losses decreased by an average magnitude of 40% compared to results obtained using RANS formulation in the aforementioned range of Re. Further, a maximum magnitude of 64% improvement in the thermal enhancement factor is achieved using LES for a B/H ratio of 0.75. Two correlations are proposed to calculate the friction factor and thermal enhancement factor for a given Re, $Nu_m$ and (B/H) ratio, with a R-squared value of 0.94 and 0.96 respectively based on the obtained simulated results The present study proposes several optimal parameters for enhancement of heat transfer in practical thermal devices.*

**Nomenclature**

| | | |
|---|---|---|
| $D_h$ | = | Hydraulic diameter of channel=$2H$ |
| $H$ | = | Distance between grooves |
| $f$ | = | Friction Factor |

---

[1] MS Scholar, Department of Applied Mechanics, IIT Madras.
[2, *] Corresponding author, MS Scholar, Department of Mechanical Engineering, IIT Madras

| | | |
|---|---|---|
| $h$ | = | Convective heat transfer coefficient |
| $k$ | = | Thermal conductivity |
| $Nu$ | = | Nusselt Number |
| $p$ | = | Static Pressure |
| $Re$ | = | Reynolds Number |
| $T$ | = | Temparature |
| $U$ | = | Mean Velocity |
| $u'_i$ | = | Fluctuation Velocity components |
| $\mu$ | = | Dynamic Viscosity |
| $\mu_{eff}$ | = | Effective Viscosity |
| $\omega$ | = | Turbulent specific dissipation rate |
| $\varepsilon$ | = | Turbulent dissipation rate |
| $\rho$ | = | Density |
| LES | = | Large Eddy Simulation |
| RANS | = | Reynold-Averaged Navier-Stokes |
| $\nabla$ | = | Spatial Dérivative |

## 1. INTRODUCTION

Heat transfer enhancement is required in nearly all heat transfer devices, including heat exchangers, biomedical applications, electronic cooling, solar air heaters, nuclear reactor cores, and gas turbine blade cooling. [1]. Several techniques are used to improve the convective heat transfer coefficient to improve overall thermal performance of these devices. Arrays of various shaped protrusions, swirl and tumble chambers, pin fins, dimple surfaces, introducing surface roughness, and rib turbulators are among the most common ways to improve heat transfer [1, 2]. All of these methods eventually result in stream-wise coherent vortices that cause advection to grow in the boundary layer next to the solid surface. The turbulence level on the boundary layer is raised by these stream-wise vortices and secondary flows, resulting in flow separation. Detached flows then pass over the next rib, forming a boundary layer, and the cycle continues. This is referred to as reattached boundary layers (hydrodynamic and thermal), which improve mixing and thus the total heat transfer coefficient and skin friction coefficient. [3]. The natural occurrence of dermal denticles on shark skins, as well as their mobility beneath the sea surface, inspired this notion [4]. However, the proper heat removal is vital to reduce heat load from the turbine aero foil blades in hot gas side using high speed air circulated for internal cooling purpose as shown in Figure 1 [5]. So, many researchers in the past studied the effect of these riblets on heat transfer augmentation. Donne and Meyer [6] conducted comprehensive measurements of heat transmission and pressure drop coefficients for various rib configurations. They compared the performance of four distinct types of artificially roughened surfaces in terms of heat transfer and friction factor. A study was carried out to determine the best artificially roughened surface for practical use. Several research on heat transfer and fluid flow have been published, using very few geometric parameter changes to keep the design simple while maintaining optimal heat transmission.

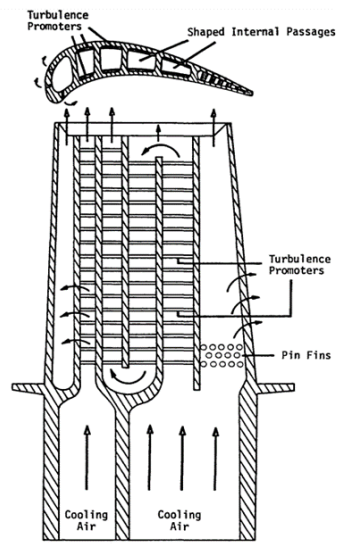

**Fig 1: Schematic design of an internally cooled gas turbine blade [5].**

In an asymmetrically channel with square ribs at varied pitch-to-height ratios, Liou and Hwang [7] employed holographic interferometry to quantify heat transfer coefficients locally while also capturing frictional loss. Within an asymmetric grooved channel with ribs for turbulent flow and heat transfer, Lorenz et al. [8] studied the distribution patterns of coefficient of heat transfer and frictional losses on the walls. They studied the local development of thermal boundary layers from a stagnation point near the front edge of a rib to the top of the rib and the front flank, comparing the average heat transfer coefficients to a plane channel. Yang and Hwang [9] studied the thermal properties of rectangular ducts with slits and ribs on one side of the wall, as well as the thermal properties of a periodic ribbed duct flow. They discovered a pair of anticlockwise whirling vortices. For a laminar incompressible flow, Mukherjee and Saha [10] investigated the influence of fillets in contracted and enlarged rectangular grooves. They used fillets of arbitrary radius to demonstrate an improvement in heat transfer. The study, however, was limited to a low Reynolds number flow, and the effect of the fillet radius on flow mechanics was not discussed. In a rectangular grooved channel, Eiamsard et al. [11] investigated the influence of different RANS turbulence models and the groove width to channel ratio. RANS models all turbulence effects while computing statistically time averaged flow quantities, resulting in very accurate forecasts and success in a wide range of industrial applications at low to moderate flow speeds. However, RANS models have an inherent drawback in modeling and its nature of making statistical averaging. So, in a high Reynolds number flows, it eventually fails to capture fine vortices and thereby miss the important flow physics [12]. On the other hand, DNS can resolve eddies of all length scales in a turbulent flow without modelling anything but it requires a very fine resolution in space and time, as well as high order differencing schemes. As a result, DNS is accurate but computationally expensive. This restricts the use of DNS in low Reynolds number flows [13]. So, eventually a robust turbulence modeling approaching is in demand that could serve both requirements: it should be accurate in resolving eddies up to fairly smaller length scales while it should be computationally less expensive. These makes it worthwhile to consider large eddy simulations (LES) to simulate a high-speed turbulent forced convective flow through a one-sided expanded grooved channel which hasn't been reported earlier. The LES solver resolves the large eddies and models the smaller eddies based on the suitable filter selection [14]. A simplified summary of the key literature reviews of the present study is illustrated in Table 1.

The objective of the present study is twofold. Firstly, the turbulent force convection is simulated for $90^0$ ribs channel using two different turbulence modelling approaches, namely, RANS and LES and the results are compared with the existing literature. Further, the supremacy of LES over RANS is well established in the present study which is first time reported in the literature till date. The augmentation in heat transfer and reduction in skin friction is reported with respect to a smooth (without ribs) channel.

So, the present paper is organized as follows: section 1 reports the introduction on the topic while section 2 describes the problem formulation, computational domain, grid convergence and a bit more detail on the RANS and LES modelling approaches. Section 3 is about the in-depth explanations of the results and finally the key findings are discussed in conclusion.

**Table 1**: Simplified tabular form of literature review

| Sr. No. | Authors | Technique used | Domain Specification | Output | Results | Identified Gap |
|---|---|---|---|---|---|---|
| 1 | Liu et al.[2] | Transient liquid-crystal technique to investigate thermal distribution in a square channel in $Re$ 15000-35000. | Turbulence promoters with different angular and V-shaped configurations. | Alteration in heat transfer distribution and friction factor with adding grooves between ribs. | Increment of 40% is achieved in the heat transfer coefficient with increase in frictional losses by less than 30% | Investigation of $Nu_m$ distribution and $C_f$ drag along the wall was not investigated. |
| 2 | Donne and Meyer[6] | Annulus experiments to develop a method to | Rods with different roughness made of | Heat transfer and pressure loss | Novel methods were developed | No use of extended surfaces with the goal of |

| | | obtain data in nuclear reactor fuel element applications. | rectangular ribs. | behaviour from rough surfaces. | to transform obtained experimental results with single rods to apply in nuclear fuel reactors. | heat transfer enhancement. Skin friction drag and shear stress due to roughed surfaces were not considered. |
|---|---|---|---|---|---|---|
| 3 | Lorenz et al. [8] | Experimental study on heat transfer and pressure drop variation along the wall of a grooved channel of turbulent flow at $10^4 \leq Re \leq 10^5$. | Insulated upper wall and a constant heat flux on lower wall using heating foils. | Temperature distributions at the walls are obtained through infrared thermography. | Existence of secondary vortex in the corner and bottom of groove, | Effect of groove width to channel height ratio on heat transfer distribution were not considered, |
| 4 | Yang and Hwang [9] | CFD analysis of heat transmission and pressure drop characteristics in rectangular duct with slits at $8000 \leq Re \leq 58000$. | Analysis at a fixed rib height to duct hydraulic diameter at different rib void fractions. | Heat transfer enhancement were investigated in detail | Occurrence of thermal enhancement and lower frictional losses due to flow pattern around ribs. | Effect of different turbulence models and varying height to duct ratio was not studied. |
| 5 | Mukherjee and Saha [10] | Numerical analysis on effect of fillets on thermo-fluid behaviour in a grooved channel at $50 \leq Re \leq 250$. | Channel consisting of 33 rectangular grooves with and without fillets | Effect of fillets on heat transfer and pressure drop were investigated | Enhancement in heat transfer due to introduction of fillets at sharp corners. | Analysis was done for a laminar flow without considering effect of fillet radius. |
| 6 | Eiamsard and | 2D Numerical analysis on heat | Consisting of 9 rectangular | Effect of turbulence model and | A maximum heat | Study was completely restricted to |

|   | Promvonge [11] | transmission over periodic grooves at $6000 \leq Re \leq 18000$ | grooves on lower side | groove width to depth ratio was investigated. | transfer enhancement factor of 1.33 is achieved using a B/H ratio of 0.75 | turbulence modelling use the RANS framework. |
|---|---|---|---|---|---|---|
| 7 | Chaube et al. [15] | CFD analysis due to roughness in form of ribs inside rectangular duct at $3000 \leq Re \leq 20000$. | Domain consists of nine multiple shaped ribs with uniform heat flux on ribbed surface. | Analysis was done to investigate heat transfer augmentation due to the rough surface using $k - \omega$ SST model. | Maximum heat transfer is obtained with chamfered ribs. | Study was restricted to RANS model and a upper $Re$ of 20000. |

## 2. METHODOLOGY

### 2.1 Description of the Computational Domain

The computational domain of the present investigation consists of a grooved channel in the horizontal plane. Figure 1(a-b) illustrates a schematic of the 3D and sectional 2D view of the grooved channel . A schematic of a single groove with the associated dimensions in shown in Figure 2. The grooves are located on the lower region of the domain. The geometric dimensions are same as mentioned in Eiamsa-ard and Promvonge [11], and the height of the expanded region H, is 40 mm.

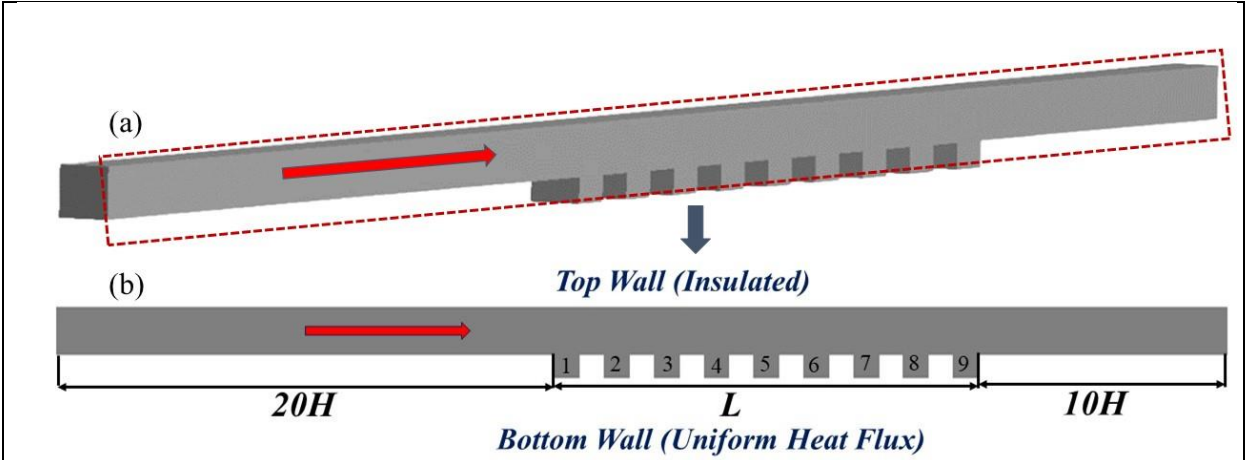

Figure 1: Schematic of the computational domain: (a) 3D view and (b) Sectional 2D view.

The upper wall is kept insulated while a constant heat flux boundary condition is applied in the lower wall as stated by Lorenz et al. [8]. The reason behind this is that the internal convective heat transfer has been into account by providing a uniform heat flux at the lower wall and neglecting any other kind of heat loss to the external environment hence, the upper wall is assumed to be insulated. Air is used as the working fluid and it travels unidirectionally along the specified direction (in positive X-axis) as shown in Figure 1. The flow is assumed to be incompressible [11]. The flow occurs over a Reynolds number of 3000 to 30,000. Pressure gradient is absent at the exit, with no slip boundary conditions at the wall [11]. Chaube et al. [15] reported that the results obtained using a 2D flow model are in good agreement with a 3D model. Further, this is verified in the present study. Hence, numerical simulations are performed using a 2D computational domain in the present work to save more computer memory, processing time, and the computational costs.

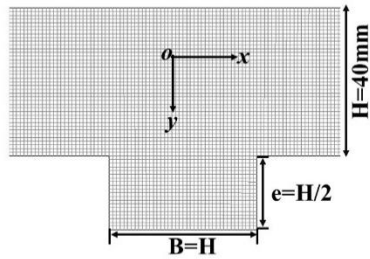

**Fig 2: Illustration of a single groove.**

The boundary conditions used in the present study are listed in the simplified Table 2 as depicted below:

**Table 2**: Details of the boundary conditions of the computational domain

| Location | Type of Boundary | In mathematical form | Magnitude | Unit |
|---|---|---|---|---|
| **Inlet** | Constant Uniform Velocity | $U = $ constant | 1.19-7 | $m/s$ |
| **Outlet** | Constant Pressure | $P = 0$ | 0 | $Pa$ |
| **Upper Wall** | Insulated | $q = 0$ | 0 | $W/m^2$ |
| **Lower Wall** | Constant Heat Flux | $q = $ constant | 2000 | $W/m^2$ |

The numerical method undertaken in the present study in based on finite volume based commercial package ANSYS Fluent 19.2

## 2.2 Details of the governing equations

The dynamics of fluid transport and thermal transmission through the flow domain are controlled by the following 2D steady equations of mass balance, momentum balance, and equation for energy transport:

*Continuity:*

$$\frac{d}{dx_i}(\rho u_i) = 0 \qquad (1)$$

*Momentum:*

$$\frac{\partial(\rho u_i u_j)}{\partial x_j} = -\frac{\partial p}{\partial x_i} + \frac{\partial}{\partial x_j}\left[\mu\left(\frac{\partial u_i}{\partial x_j} + \frac{\partial u_j}{\partial x_i}\right)\right] + \frac{\partial}{\partial x_j}(-\rho\overline{u'_i u'_j}) \qquad (2)$$

*Energy:*

$$\frac{\partial}{\partial x_i}(\rho u_i T) = \frac{\partial}{\partial x_j}\left((\Gamma + \Gamma_t)\frac{\partial T}{\partial x_j}\right) \qquad (3)$$

In expression (3), $\Gamma$ and $\Gamma_t$ represents the two types of thermal diffusivity: molecular and turbulent respectively.

### 2.3 Turbulence Modelling

#### 2.3.1 Reynolds averaged Navier-Stokes (RANS) framework

RANS is the most affordable and widely used modelling approach in the majority of industrial scale turbulent flow scenarios where a fair match between average accuracy and computation cost is uttermost crucial. The mass, momentum, and energy equations accept the decomposed mean and fluctuation components of the dependent variables from RANS. To create equations for mean variables, these data are also averaged across time. The effects of variable quantities on the mean flow must be modelled in order to provide a closure for mean flow equations. More information on RANS is provided elsewhere [16–19]. However, the RNG k-turbulence model is used in this work.

The governing equations for the RNG k-$\epsilon$ model are:

$$\frac{\partial}{\partial t}(\rho k) + \frac{\partial}{\partial x_i}(\rho k u_i) = \frac{\partial}{\partial x_j}\left(\alpha_k \mu_{eff} \frac{\partial k}{\partial x_j}\right) + G_k + G_b + \rho\varepsilon - Y_M + S_k \qquad (4)$$

$$\frac{\partial}{\partial t}(\rho\varepsilon) + \frac{\partial}{\partial x_i}(\rho\varepsilon u_i) = \frac{\partial}{\partial x_j}\left(\alpha_\varepsilon \mu_{eff}\frac{\partial \epsilon}{\partial x_j}\right) + C_{1\varepsilon}\frac{\varepsilon}{k}(G_k + C_{3\varepsilon}G_b) - C_{2\varepsilon}\rho\frac{\epsilon^2}{k} - R_\varepsilon + S_\varepsilon \qquad (5)$$

Mean velocity gradients that cause turbulence by creating kinetic energy is denoted by $G_k$ in the equations 4&5 while the production of kinetic energy during turbulence because of buoyancy is given by $G_b$. Variable dilatation in compressible turbulence and its impact on a measure of overall dissipation is represented by $Y_M$. For $k$ and $\epsilon$ the reverse $Pr$, are $\alpha_k$ and $\alpha_\varepsilon$ respectively. $S_k$ and $S_\varepsilon$ represent the individual source terms.

The RANS approach to create a turbulence model requires the Reynolds stress term denoted by $-\rho \overline{u'_i u'_j}$ in equation (2) to be modelled. We choose the $k - \varepsilon$ turbulence framework for closure of the equations. The Reynolds stress term and the mean velocity gradients term are related using the widely used Boussinesq technique. It is given by:

$$-\rho \overline{u'_i u'_j} = \mu_t \left( \frac{\partial u_i}{\partial x_j} + \frac{\partial u_j}{\partial x_i} \right) \quad (6)$$

In the above expression (6), $\mu_t$ denotes the term which computes the turbulent viscosity. It is given by:

$$\mu_t = \rho C_\mu \frac{k^2}{\varepsilon} \quad (7)$$

The expression for the TKE production, $k$ is written as:

$$\frac{\partial}{\partial x_i}(\rho \varepsilon u_i) = \frac{\partial}{\partial x_j}\left[\left(\mu + \frac{\mu_t}{\sigma_k}\right)\frac{\partial k}{\partial x_j}\right] + G_k - \rho \varepsilon \quad (8)$$

The corresponding equation for the turbulence destruction of the TKE, $\varepsilon$ is shown below:

$$\frac{\partial}{\partial x_i}(\rho \varepsilon u_i) = \frac{\partial}{\partial x_j}\left[\left(\mu + \frac{\mu_t}{\sigma_\varepsilon}\right)\frac{\partial \varepsilon}{\partial x_j}\right] + C_{1\varepsilon}\frac{\varepsilon}{k}G_k - C_{2\varepsilon}\rho\frac{\varepsilon^2}{k} \quad (9)$$

In the above expression (9), $G_k$ denotes the rate of energy production due to turbulence and $\rho \varepsilon$ denotes the rate of destruction. $G_k$ is further denoted by:

$$G_k = -\rho \overline{u'_i u'_j} \frac{\partial u_j}{\partial x_i} \quad (10)$$

We define and specify the boundary value of the turbulent quantities proximal to the wall using the enhanced wall treatment method in ANSYS Fluent 19.2®. The empirical constants are taken as $C_\mu$=0.087, $C_{1\varepsilon}$=1.43, $C_{2\varepsilon}$=1.9, $\sigma_\varepsilon$=1.32 and $\sigma_k$=0.98 respectively in the governing transport equation of turbulent flow.

### 2.3.2 Large Eddy Simulations (LES) framework

LES is among the most accurate modelling approaching techniques available in the literature after DNS. It is computationally expensive than RANS. The mass, momentum, and energy balance equations are represented in LES using low-pass filtering the mass, momentum, and energy equations of small regions in physical space of characteristic size Δ. This results in a coarse-grained physical representation of the flow, with the time dependent terms of the resolved scales, which is directly controlled by the boundary conditions, preserved, but the influence of small scale, subgrid, turbulence must be provided by a model. The details on the LES could be found in literature [14] in more details. The dependent variables are split into resolved and subgrid components in LES, and then inserted into the mass, momentum, and energy balance equations, which are further filtered to yield transport equations for the resolved quantities.

$$\frac{\partial \bar{u}_i}{\partial x_i} = 0, \tag{6}$$

$$\rho \frac{\partial \bar{u}_i}{\partial t} + \rho_0 \frac{\partial (\bar{u}_i \bar{u}_j)}{\partial x_j} = -\frac{\partial \bar{p}^*}{\partial x_i} + \frac{\partial}{\partial x_j}\left(2\rho_0 v \bar{S}_{ij} - \tau_{ij}^{SGS}\right) - \rho_0 \beta (T - T_0) g_i \tag{7}$$

$$\rho_0 \frac{\partial \bar{e}}{\partial t} + \rho_0 \frac{\partial (\bar{u}_j \bar{e})}{\partial x_j} = -\bar{p}\frac{\partial \bar{u}_i}{\partial x_i} + 2\rho_0 v \bar{S}_{ij}\frac{\partial \bar{u}_i}{\partial x_j} + \frac{\partial}{\partial x_i}\left(\bar{h}_j - q_j^{SGS}\right) \tag{8}$$

In the expression 6, 7 and 8, the filtered variables are denoted with an overbar over them. $\tau_{ij}^{SGS}$ and $q_j^{SGS}$ are the subgrid stress tensor and flux vector respectively. Although the Smagorinsky-Lilly model is widely used in LES simulations but it has its own limitations. The Smagorinsky model is a dissipative eddy-viscosity model that cannot forecast energy transmission from sub-scale grid structures to larger resolution scales. Further this model usually fails to give a non-zero viscosity near the wall and usually fails to incorporate the effects of shear and rotation in the near wall zone. In the present study, we use WMLES, a wall modelled LES approach, where a sub grid model is applied to close the LES equation and model the flux vector and sub grid scale stress tensor. In the near wall regions, a different model is used to represent the closeness to the wall zone. Based on literature, it can be concluded that $\tau_{ij}^{SGS}$ can be modelled by a transport equation. It is further represented as:

$$\tau_{ij}^{SGS} = -2\rho_0 v_{SGS}\bar{S}_{ij} + \frac{2}{3}\rho_0 k \delta_{ij} \tag{9}$$

In expression (9), $\bar{S}_{ij}$ is the resolved rate of strain tensor and $v_{SGS}$ is the subgrid viscosity. $\tau_{ij}^{SGS}$ is modelled using the wall adaptive local eddy viscosity (WALE) model and $q_j^{SGS}$ is modelled using gradient hypothesis. In expression (9), modelling of the term $v_{SGS}$ term is done in the following manner as shown below in expression (10).

$$v_{SGS} = (C_W \Delta)^2 \frac{\left(\mathfrak{H}_{ij}^d \mathfrak{H}_{ij}^d\right)^{3/2}}{\left(\bar{S}_{ij}\bar{S}_{ij}\right)^{5/2} + \left(\mathfrak{H}_{ij}^d \mathfrak{H}_{ij}^d\right)^{5/4}} \tag{10}$$

In the above expression, $\Delta$ represents the cell size and $C_W$ represents a model parameter whose magnitude roughly equates to 0.33. $\mathfrak{H}_{ij}^d$ is defined as shown below in expression (11).

$$\mathfrak{H}_{ij}^d = \frac{1}{2}\left(\bar{g}_{ij}^2 + \bar{g}_{ji}^2\right) - \frac{1}{3}\delta_{ij}\bar{g}_{kk}^2 \tag{11}$$

In the above expression (11), $\bar{g}_{ij} = \frac{\partial \bar{u}_i}{\partial x_j}$ and $\bar{g}_{ij}^2 = \bar{g}_{ik}\bar{g}_{kj}$. The $q_j^{SGS}$ term in expression (8) is denoted by:

$$q_j^{SGS} = -\rho_0 \alpha_{SGS} \frac{\partial \bar{e}}{\partial x_j} \qquad (12)$$

Further, $\alpha_{SGS} \approx \frac{\nu_{SGS}}{Pr}$, where $Pr$ denotes the sub grid Prandtl number whose magnitude is roughly taken as 0.7. The prime objective of using WMLES is to resolve the large eddies having higher magnitude of energy associated with them and to model the smaller eddies.

### 2.4 Domain discretization and mesh sensitivity analysis

For any numerical computation using CFD, finding the optimal is an integral part and based on that, an optimum grid is chosen for the flow simulation. Figure 3 depicts the variation of co-efficient of heat transfer at the bottom wall over five different mesh sizes. The laminar sub-layer (or, viscous sub layer in many references) is resolved in rectangular grids using face meshing techniques, with grid adoption for y+ ≈2 at a near wall zone [11]. To establish the optimal mesh size for numerical simulations, a grid sensitivity analysis is further carried out. The number of cells used to attain grid independence is varied between 50,900 and 155,992 rectangular types of meshes and the heat transfer coefficient is calculated for each case at a Reynolds number of 12000. The optimum grid is obtained, keeping residual less than $10^{-3}$. LES simulations were performed on the optimum grid obtained from RANS framework. However, the mesh in adjusted and made finer on the individual cases depending upon the varied Reynolds number at the inlet to effectively capture the eddies in the vicinity of the wall of the channel.

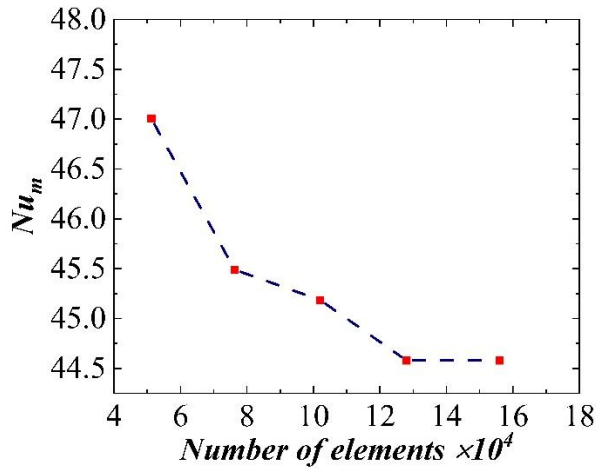
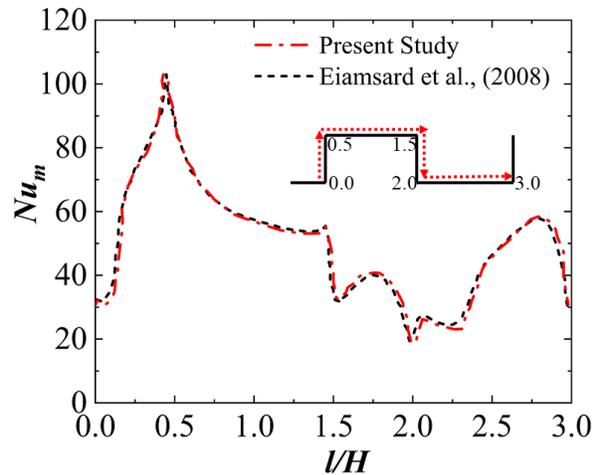

Figure 2: Mesh sensitivity analysis for $Nu_m$ over different grid sizes.

Figure 3: Comparative analysis for a mesh size of 155,992 of variation $Nu_x$ along the groove wall at Re=12000 and B/H=1.0.

### 2.5 Model Validation

The results obtained through numerical simulations were validated for a steady incompressible flow at a *Re* of 12000 using experimental results of Lorenz et al. [8] and Eiamsa-ard et al. [11] respectively. The variation of the local Nusselt number ($Nu_x$) is investigated along the wall of a single groove as shown in Figure 4 for a groove width to depth ratio (B/H) of 1. The results of the present simulation are observed to satisfy a reasonable match with literatures.

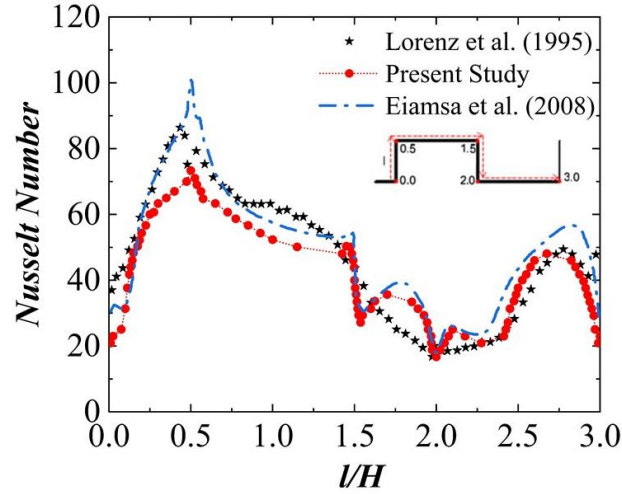

Figure 4: Comparison of local Nusselt number ($Nu_x$) distribution along the groove wall with literature [8,11].

### 2.6 Numerical Methodology

The numerical simulations in the present study are based on the grid centered discretization using the finite volume method. To attain second order spatial precision, convective fluxes are modelled using central differencing, whereas diffusive fluxes are reconstructed using a fully conservative central differencing technique. The second order accurate implicit method is used for time stepping, with a maximum Courant–Friedrichs–Lewy (CFL) number of 1 for numerical stability. The semi-implicit method for pressure linked equations (SIMPLE) algorithm is used for the pressure-velocity coupling. Bounded central differencing technique is used to discretize the momentum in LES framework. Second order upwind method is used to model the energy with a second order transient implicit formulation. The parameters of interest for present study are (1) Nusselt Number, (2) Frictional losses, (3) Skin Friction Coefficient, and (4) Thermal Enhancement Factor respectively. We compute the friction factor value (*f*) using the pressure losses ($\Delta p$) across the computational domain having the hydraulic diameter, $D_h = 2H$. It is given as:

$$f = \frac{\left(\frac{\Delta p}{L}\right) D_h}{\frac{1}{2}\rho u^2} \tag{13}$$

The coefficient of skin friction drag denoted by $C_f$, is defined as:

$$C_f = \frac{\tau_{wall}}{\frac{1}{2}\rho u^2} \tag{14}$$

The magnitude of thermal energy transferred is calculated by $Nu_m$ which is computed by the following expression:

$$Nu(x) = \frac{h(x) D_h}{k} \tag{15}$$

The mean Nusselt number is calculated by:

$$Nu = \frac{1}{L}\int Nu(x)Ax \qquad (16)$$

We calculate the thermal enhancement factor, $\eta$, using the following expression:

$$\eta = \left(\frac{Nu}{Nu_s}\right)/\left(\frac{f}{f_s}\right)^{1/3} \qquad (17)$$

### 3. Results and Discussions

### 3.1 Comparative analysis of heat transfer

In this section, we investigate the relationship and the flow physics of the associated working fluid inside the grooved channel and the heat transfer. Figure 5 illustrates a variation of the mean Nusselt number ($Nu_m$) with the Reynolds number ($Re$) modelled using LES (WMLES) framework used in the present study, compared with existing RANS results from literature of Eiamsard et al. [11] for a groove width to depth ratio (B/H) ranging from 0.75 to 1.75. It is observed that the obtained magnitude of heat transfer is relatively higher using a LES model compared to the existing RANS results. Further, the magnitude of heat transfer decreases with increase in the B/H ratio. It is to be noted that the time-averaged equation of the RANS model is based on isotropic assumptions, as a result of which it fails to capture the large-scale eddies and circulations in the flow domain which has substantial effect on the computed value of the heat transfer coefficients. However, the difference in magnitude of the heat transfer obtained using both the model decreases with increase in the mean inlet velocity for a given B/H.

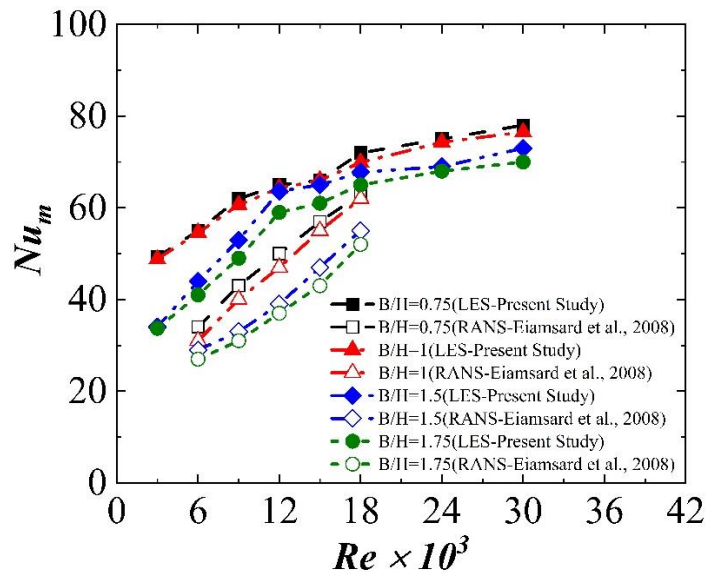

Figure 5: Comparative variation of the mean Nusselt number with $Re$ at different B/H using LES and RANS.

Literature results obtained using the RANS framework, suggests that the variation of the $Nu_m$ is approximately linear in the range of Re from 6000 to 18000. However, results from the present study obtained using the LES framework depicts a non-uniform variation of the $Nu_m$ in the range of Re from 3000 to 30,000 done in the present study. The rate of increment of the heat transfer decreases with the increase in Re or the mean inlet velocity. The heat transfer coefficient is significantly higher by a magnitude of 51% compared to reported RANS results at a $Re$ of 6000. An overall increment of 45% in the magnitude of heat transmission ($Nu_m$) is achieved using LES framework in $3000 \leq Re \leq 30000$. Figure 6(a-d) illustrates a comparative variation of the mean Nusselt number with Re at different B/H ratios. It is observed that the margin of difference between the $Nu_m$ is maximum for B/H=0.75 and minimum for B/H=1.75. In the next section, we visualize the velocity streamlines and the associated structures of the fluid flow at different ratios of B/H to capture the related flow physics responsible for the heat transfer enhancement.

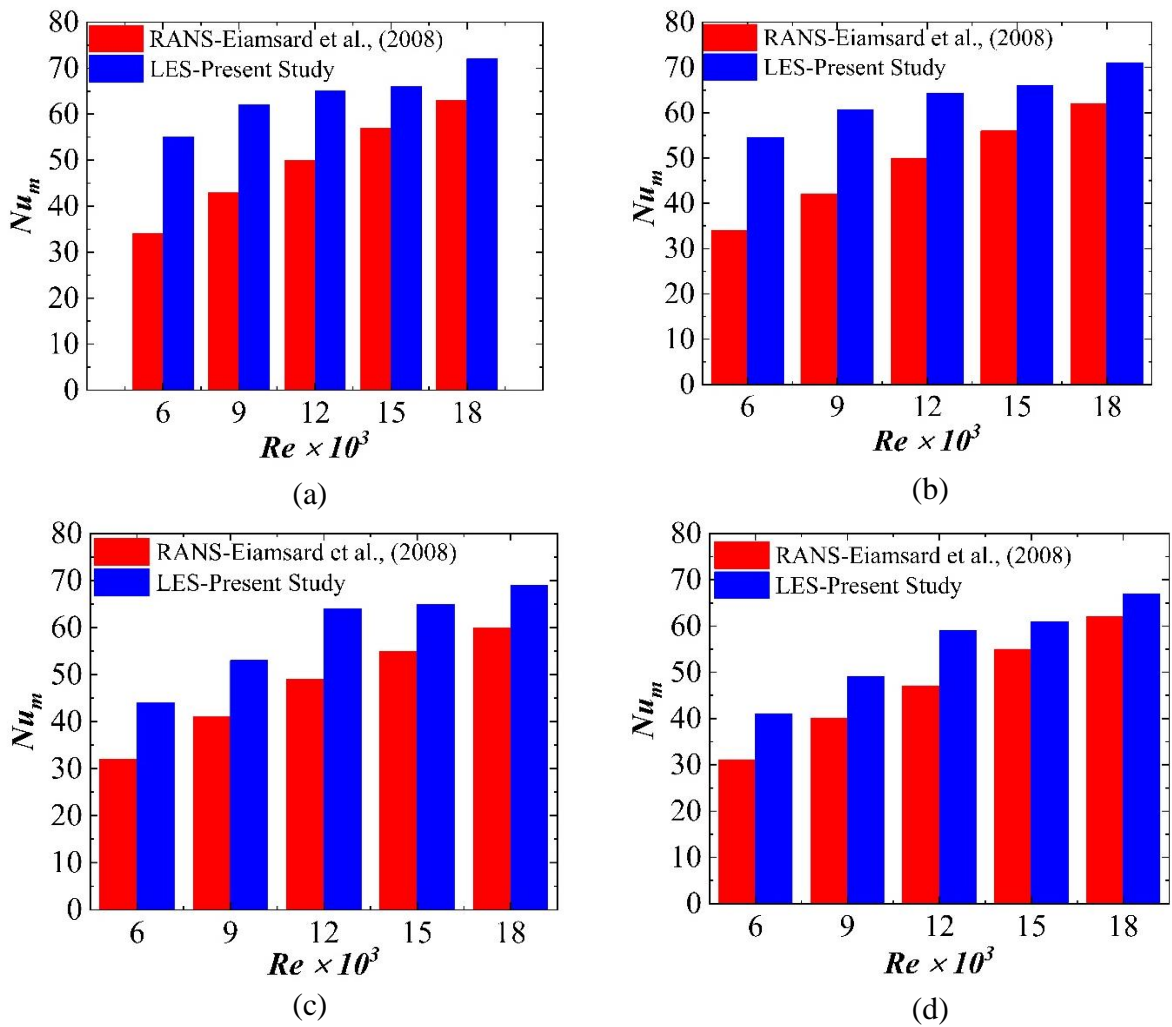

Figure 6: Comparative histogram of variation of $Nu_m$ with Re at different B/H ratios: (a) 0.75, (b) 1.0, (c) 1.5 and (d) 1.75 respectively.

### 3.2 Velocity Streamlines and fluid flow structure

Figure 7.1 to 7.4 illustrates the velocity streamlines inside the $6^{th}$, $7^{th}$ and the $8^{th}$ groove of the channel modelled using the LES framework for a *Re* of 3000 and 30,000 with the groove width to height (B/H) ratio ranging from 0.75 to 1.75 respectively. It gives us a general feature of the flow and the effect of the B/H ratio on the heat transfer enhancement with flow velocity. After the flow gets separated from the top of the channel, a large recirculation zone is formed in the grooved region. Multiple smaller recirculation zones can be observed in all the cases, around the bottom corners of the grooved using the LES model which the RANS framework in literature, failed to capture effectively. The formed vortices and eddies enhances the thermal transport process at the vicinity of the lower wall by creating a swirling effect and destabilizing the flow field. The flow again attaches at the top half of the channel. As evident from the figure, the fluid flow behaviour is complex and the and recirculation zones and eddies in the grooved region increases with *Re*. These eddies observed in the velocity streamlines enhances the turbulent heat transmission through strong turbulence mixing which results in subsequent transport of heat energy. Thermal energy is transmitted more directly and effectively to the surrounding fluids through the large eddies formed by the separated shear layers due to the movement of the reattachment zone which inhibits development of any significant boundary layer. There is significant rise is turbulence energy dissipation rate which increase the mixing in the vicinity of the wall region of the groove. This is primarily due to the localised turbulence energy dissipation rate which results in occurrence of high magnitude of inhomogeneity in the grooved channel generated by the eddies and the transverse vortices. Finite magnitude of velocity in the near wall region proximal to the groove increase the shear stress on the wall. The vortices result in high turbulence energy dissipation rate and thus enhanced mixing which increases the heat transfer. A net increase of heat transfer rate by a reasonable magnitude of 45% is achieved in the range of Re 3000 to 30000.

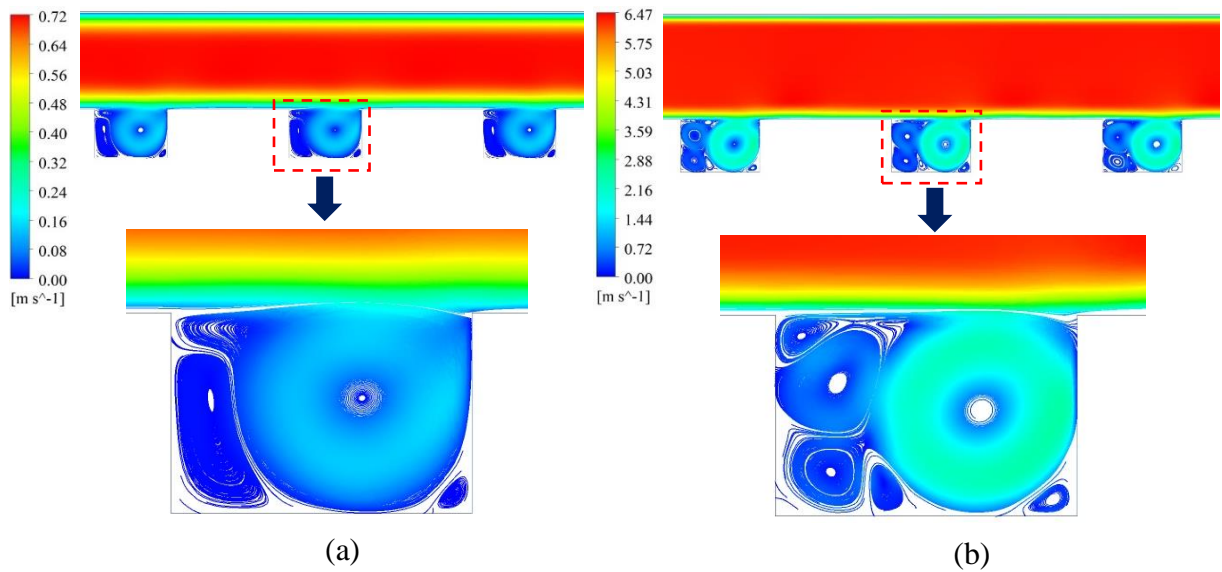

(a)                                                                                  (b)

Figure 7.1 : Velocity streamlines at B/H=0.75 at : (a) Re=3000 and (b) Re=30000 respectively.

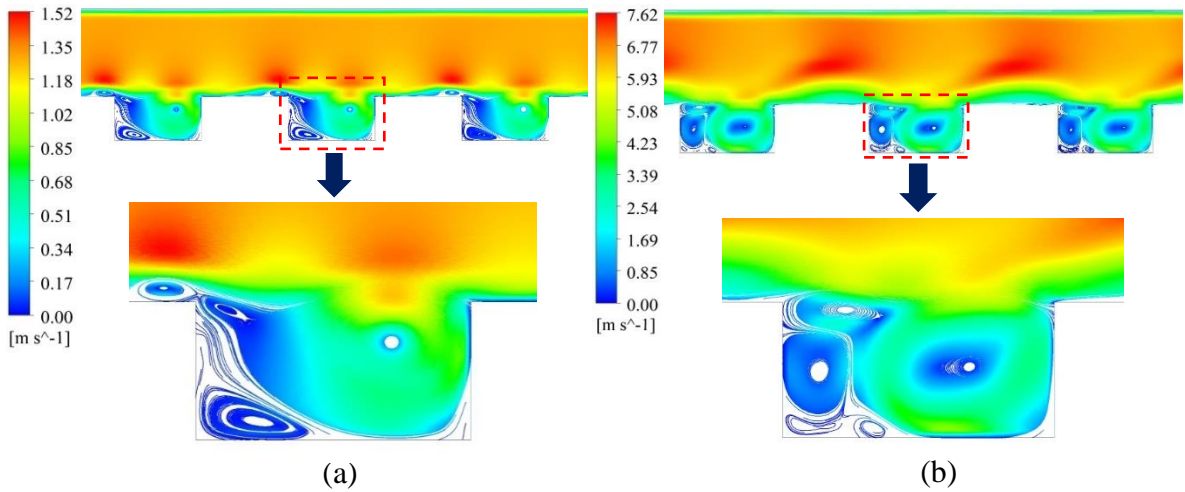

Figure 7.2 : Velocity streamlines at B/H=1 at : (a) Re=3000 and (b) Re=30000 respectively.

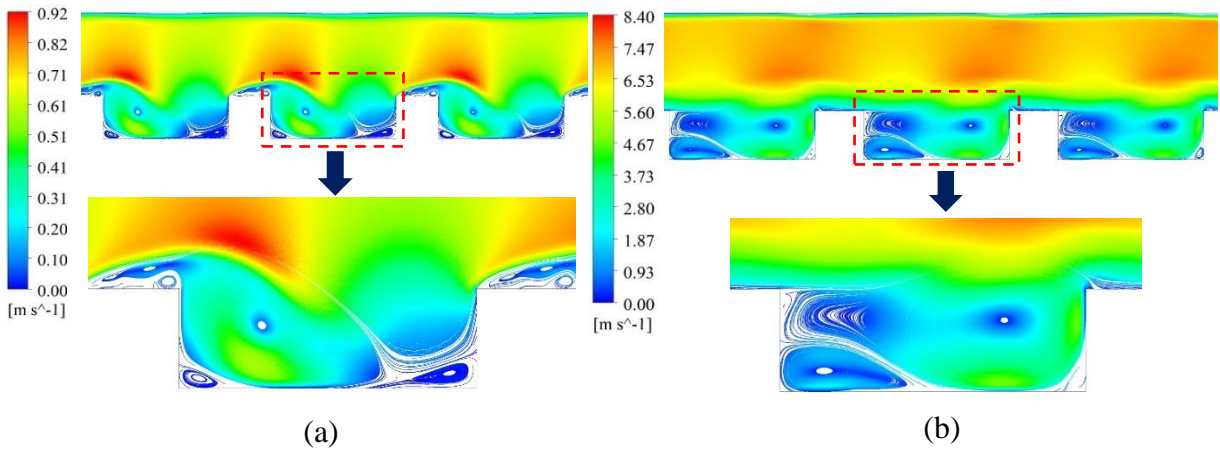

Figure 7.3 : Velocity streamlines at B/H=1.5 at : (a) Re=3000 and (b) Re=30000 respectively.

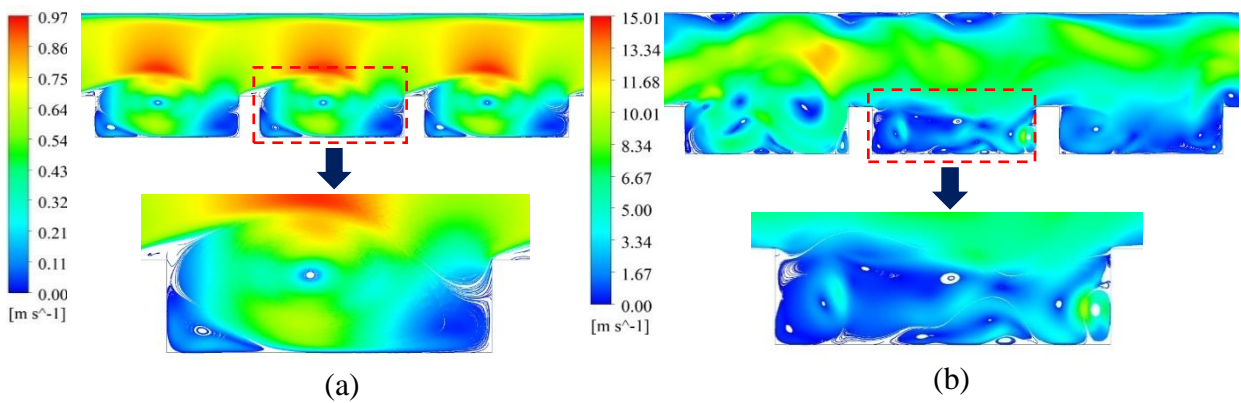

Figure 7.4 : Velocity streamlines at B/H=1.75 at : (a) Re=3000 and (b) Re=30000 respectively.

### 3.3 Effect of groove depth to width (B/H) ratio in heat transfer

A significant change in the structure of the eddies and vortices is observed in the grooved region with the increase in B/H ratio. At B/H=0.75, a large recirculation zone is observed in the 7$^{th}$ groove of the channel. The number of transverse vortices increases with increase in the *Re*. These multiple rotating structures generates instability in the flow and exchange of fluid between the wall layer and the main flow midway through the channel. This leads to enhanced mixing and energy transfer between the eddies. Four prominent rotating transverse vortices are observed at B/H=0.75 for a *Re* of 30,000. Transverse vortices contributes to negligible heat transfer enhancement in a steady flow [20]. The high speed transient nature of the flow in the present study leads to significant heat transfer improvement due to the transverse vortices. These transverse vortices have significant impact on the heat transfer by destabilizing the flow which further results in formation of self-exciting vortices with their axis perpendicular to the direction of the flow. The number of vortices reduces with increase in the B/H ratio for a given Re which further explains the decreasing trend in the heat transfer enhancement with increase in B/H.

The enhancement in the heat transfer is accompanied with a pressure loss penalty across the flow domain. In the next section, we investigate the encountered frictional losses across the flow domain for the different geometric configurations.

### 3.4 Investigation of Frictional Losses

We investigate the variation of the pressure losses across the computational domain for a range of B/H ratio from 0.75 to 1.75 in *Re* ranging from 3000 to 30000 using LES modelling. Figure 8 illustrates a variation of the friction factor with Re for different width to height ratios (B/H) in the aforementioned range of Reynolds number. The encountered frictional losses obtained using the LES simulation is lesser in magnitude compared to that obtained using the RANS framework and it decreases with increase in *Re* for a particular B/H ratio.

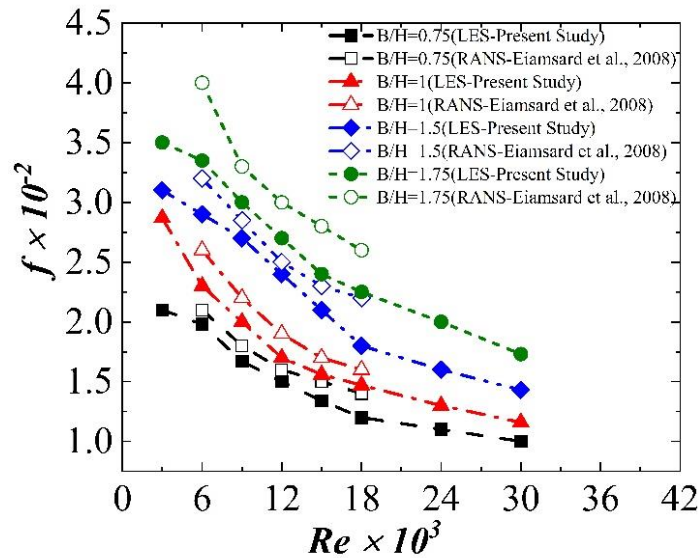

Figure 8: Comparative variation of friction factor with *Re* at different B/H using LES and RANS.

A comparative histogram of the frictional losses of the present work with reported literature of Eiamsard et al. [8]. It is observed that the computed value of the friction factor increases with the increase in B/H ratio. The B/H =1.75 shows the maximum value of the friction factor

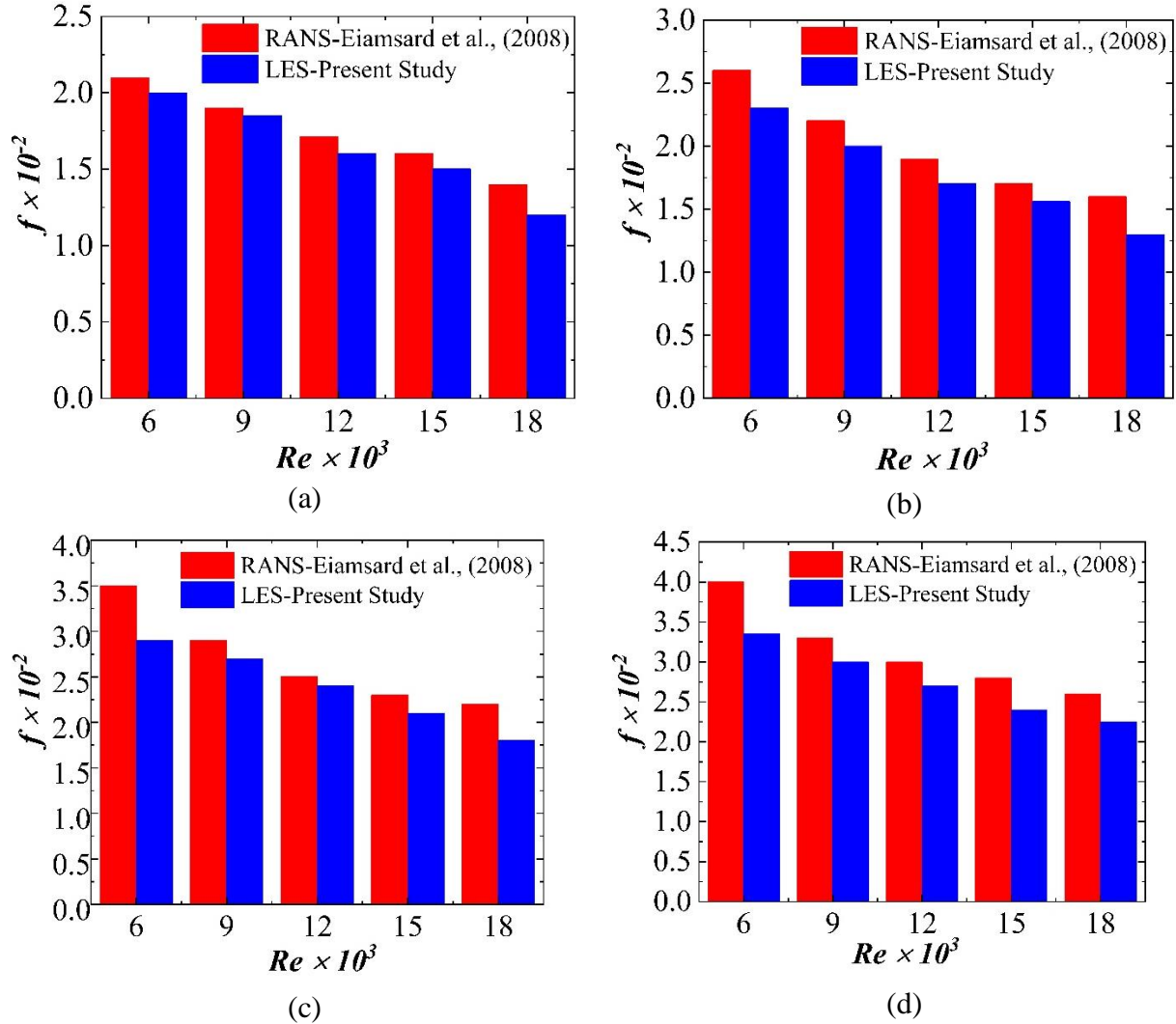

Figure 9: Comparative histogram of variation of friction factor with *Re* at different B/H ratios: (a) 0.75, (b) 1.0, (c) 1.5 and (d) 1.75 respectively.

Figure 10 depicts the distribution of the pressure contours across the 6$^{th}$, 7$^{th}$ and 8$^{th}$ groove respectively at two different *Re*. It is now well established from the analysis of section 3.2 that the grooved region, as captured using LES, forms eddies and recirculation zones. These eddies quickly arise and dissipate as a result of increased turbulence. This causes the creation of a low-pressure zone, which reduces the shear stress on the channel's wall regime. This lowers the overall pressure drop and frictional loss by an average of 40% compared to that reported in literature obtained using the RANS model, in the aforementioned range of *Re*. The RANS model fails to effectively capture these low pressure zones at varying B/H ratio and flow speeds. Further the area of the low pressure zone in the grooved region increases with the flow speed.

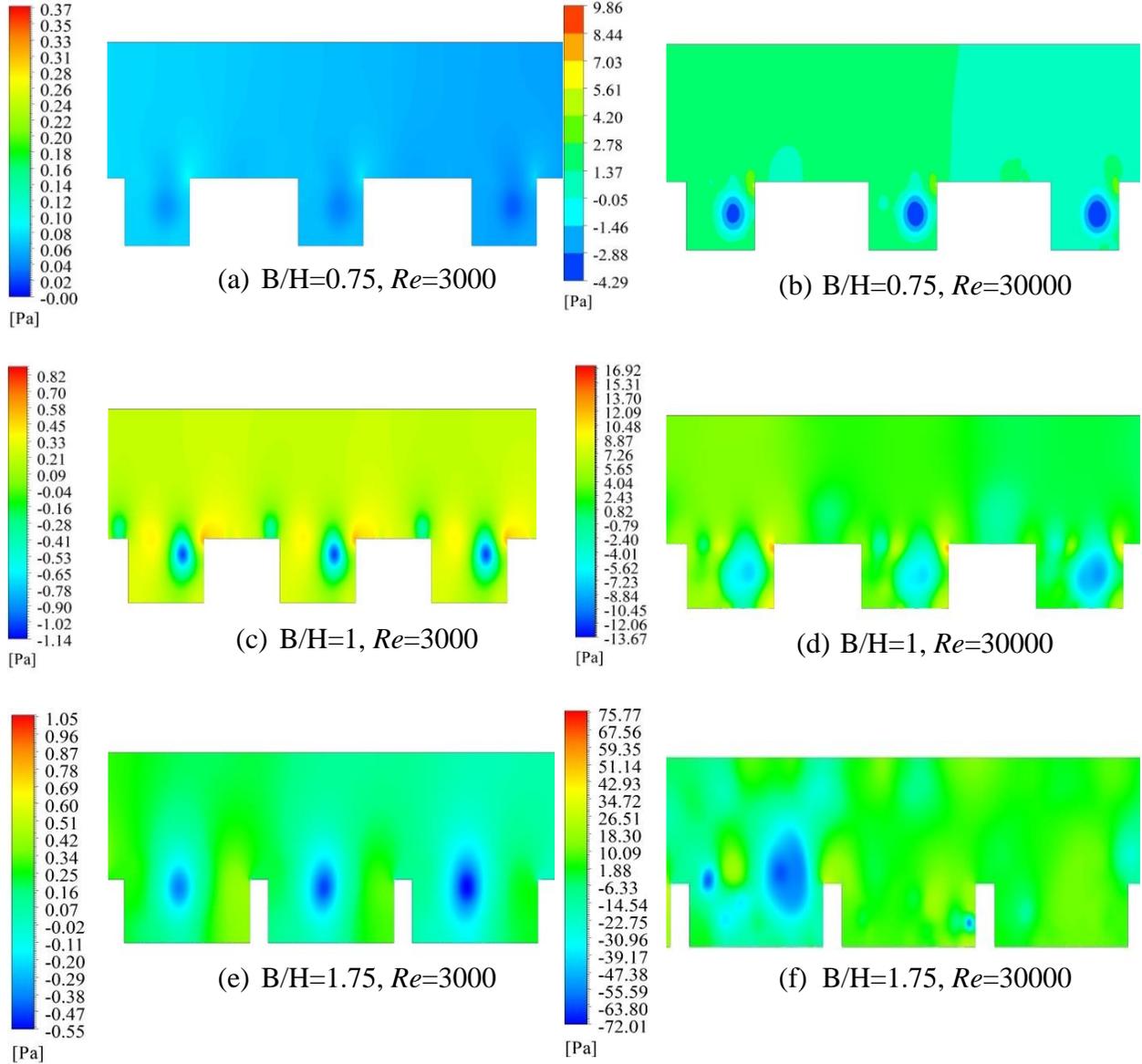

Figure 10: Distribution of pressure contours at 6$^{th}$, 7$^{th}$ and 8$^{th}$ groove respectively for different B/H ratios for *Re* of 3000 and 30000.

A correlation is proposed to calculate the friction factor for a given value of Reynolds number, Nusselt number and (B/H) ratio based on the obtained LES results with $R^2 = 0.94$, as shown below in expression 18.

$$f = 0.2 Re^{-0.49} Nu^{0.53} \left(\frac{B}{H}\right)^{0.7} \tag{18}$$

### 3.5 Temperature distribution

Figure 11 illustrates the distribution of the temperature contours across the 6$^{th}$, 7$^{th}$ and 8$^{th}$ groove respectively at two different *Re*. We observe three separate temperature sublayers extending from the groove's lower wall to the channel's top. These layers are most prominent and distinct for B/H=0.75 and gradually overlaps with increase in the B/H value. In the contours, it is observed that the temperature distribution become unstable and air of a relatively cooler temperature penetrates the grooved region of the channel. This phenomenon results in creation of high magnitude of oscillations of the thermal boundary layers.

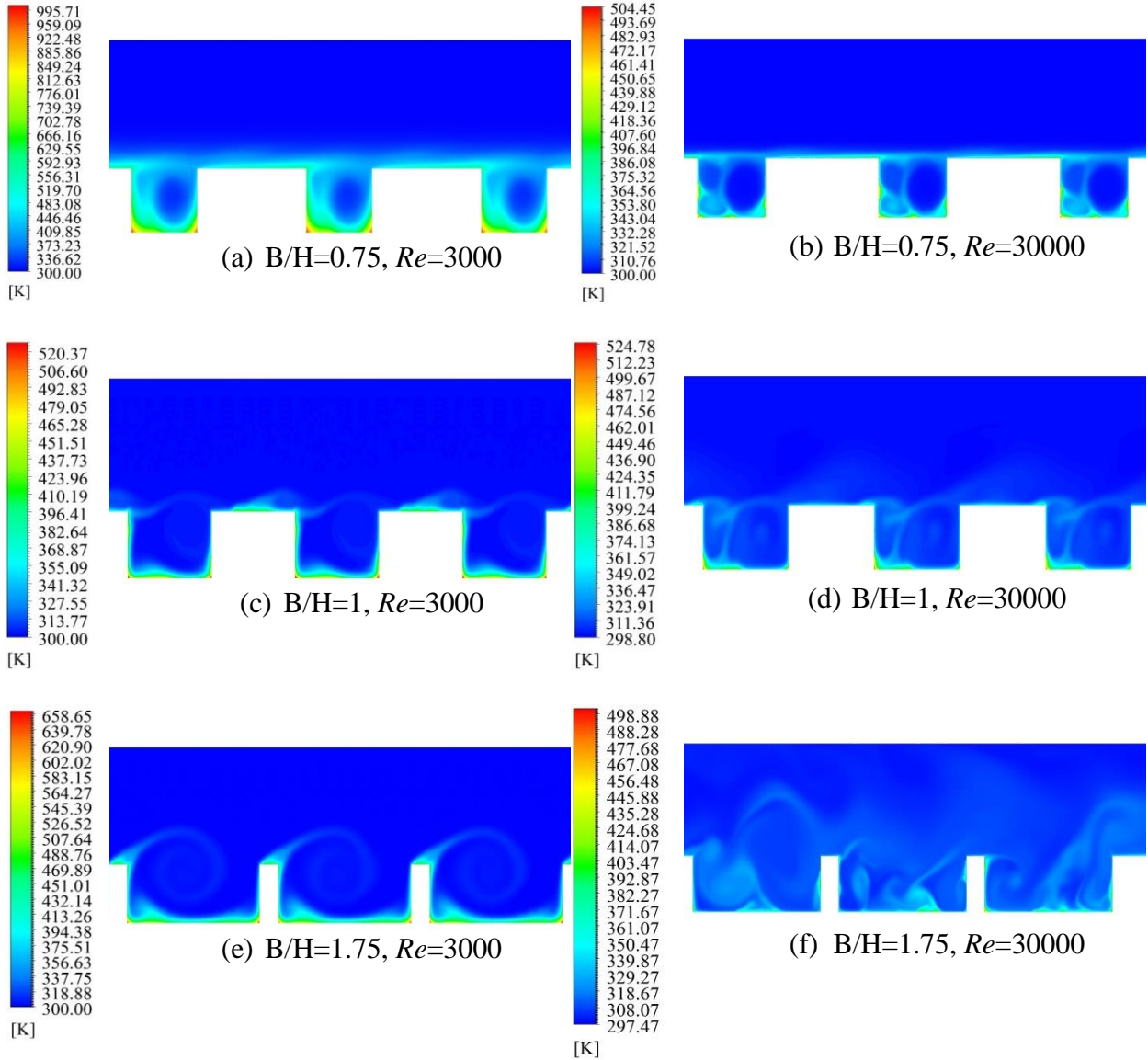

Figure 11: Distribution of temperature contours at 6$^{th}$, 7$^{th}$ and 8$^{th}$ groove respectively for different B/H ratios for *Re* of 3000 and 30000.

It is observed that the temperature of the air penetrating the grooved region is relatively hotter compared to that of the mainstream fluid. The mixing of the working fluid among the different temperature zones, which is seen to be highest for B/H=0.75 at the two *Re*, enhances the heat transmission in the grooved wall surface.

### 3.6 Evaluation of thermal enhancement factor

The thermal enhancement factor ($\eta$) in the present study is calculated using the mathematical expression as illustrated in reported literatures of Eiamasard et al. [11] as given in expression 17.

$$\eta = \frac{\left(\frac{Nu}{Nu_s}\right)}{\left(\frac{f}{f_s}\right)^{1/3}}$$

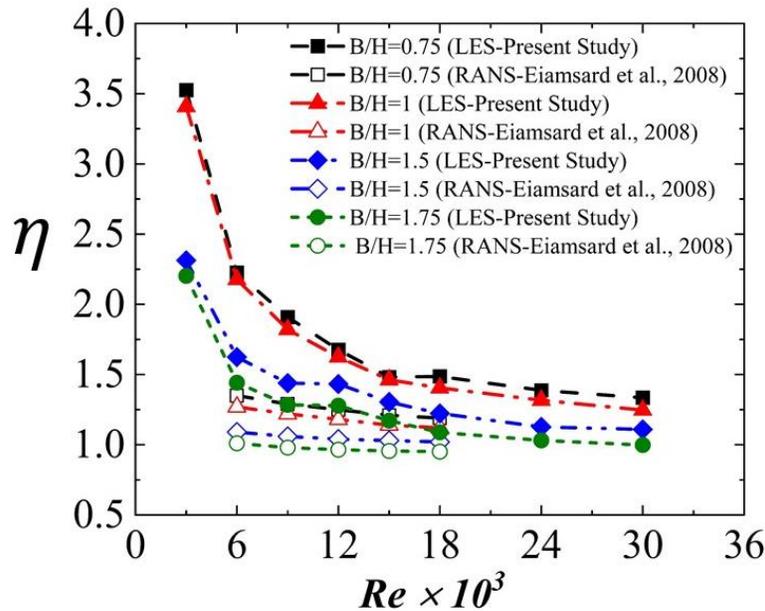

Figure 12: Comparative variation of thermal enhancement factor factor with *Re* at different B/H using LES and RANS.

Figure 12 illustrates the comparative variation of the magnitude of the thermal enhancement factor with Reynolds number for B/H ratios ranging from 0.75 to 1.75 obtained using LES framework with reported literature of Eiamsard et al.[11]. A maximum magnitude of 64% improvement in the thermal enhancement factor is achieved using LES compared to literature results using RANS for a B/H ratio of 0.75 at a Reynolds number of 3000. As discussed in section 3.4, the magnitude of computed frictional losses across the domain is significantly lesser using the LES framework compared to RANS. This further, increases the magnitude of $\eta$ for a given B/H ratio. The enhancement factor decreases with increase in the flow speed for a particular value of B/H ratio. The value of $\eta$ is highest with B/H=0.75, and lowest with B/H=1.75. The transverse nature of the grooves creates energy loss in the fluid motion. The magnitude of these losses is low at lower value of Reynolds number. We propose a correlation to calculate the $\eta$ for a given value of Reynolds number, Nusselt number and (B/H) ratio based on the obtained LES results with $R^2 = 0.96$ , as shown below in expression 19.

$$\eta = 22.28 Re^{-0.64} Nu^{0.83} \left(\frac{B}{H}\right)^{-0.21} \tag{19}$$

**Conclusions**

Understanding the flow dynamics and investigating the thermal characteristics of turbulent convective behaviours across a periodic transverse grooved channel in a reasonably high-speed flow regime using large eddy simulations is the goal of the present research. The obtained results are compared with existing RANS results in literature. Effect of the large-scale vortices and eddies captured using the LES framework on the heat transfer and frictional losses are are the key findings of the present work. In comparison to results reported using RANS formulation in literature, the results acquired using LES demonstrate improvements in the heat transfer rate by a reasonable magnitude of 45 percent while the incurred frictional losses across the flow domain dropped by an average magnitude of 40 percent in the range $3000 \leq Re \leq 30000$. Further, for a B/H ratio of 0.75, LES can improve the thermal enhancement factor by a maximum magnitude of 64 percent. Two correlations are proposed with an R-squared value of 0.94 and 0.96, respectively, to determine the friction factor and thermal enhancement factor for a given *Re*, *Nu*, and (B/H) ratio. The present work shows a significant impact of the near wall large eddies and vortices on the net thermal performances of these grooved channels at different groove width to height configuration, and captures the governing flow physics effectively using a computational approach.

**Acknowledgements**

The authors are thankful to the PG Senapathy Computing Resources at IIT Madras. The authors sincerely acknowledge IIT Madras for providing a quality education based on which they are able to pursue the present research independently.